# General-Purpose Photonic Computing Primitive for Contemporary Artificial Intelligence

Shupeng Ning,[a] Chenghao Feng,[a] Zhenxiang Xu,[a] Hanqing Zhu,[a]
David Z. Pan,[a] Jiaqi Gu,[b] and Ray T. Chen[a,*]

[a] *Department of Electrical and Computer Engineering, The University of Texas at Austin, Austin, Texas 78758, United States*

[b] *School of Electrical, Computer and Energy Engineering, Arizona State University, Tempe, Arizona 85281, United States*

[*] *Email: chenrt@austin.utexas.edu*

## Abstract

Photonic computing offers a promising route to accelerating artificial intelligence (AI) by providing high analog bandwidth, low latency, and low energy consumption. However, existing optical neural networks (ONNs) struggle with substantial hardware overhead and limited support for the dynamic, arbitrary matrix operations essential for modern AI architectures. Here we present the dynamic universal encoding tensorcore (DUET), a general-purpose photonic computing paradigm based on vectorized operand differential interferometric cells (VODICs). By exploiting inherent structural symmetry, this design provides a full-range linear encoding interface that directly accommodates signed operands. This approach eliminates the sign-based path splitting, nonlinear remapping, and auxiliary preprocessing typically required in conventional ONNs, thereby reducing latency and minimizing hardware and memory overhead. We further implement a hardware-aware training (HAT) strategy to alleviate the impact of on-chip non-idealities and ensure stable inference. DUET is experimentally validated across diverse architectures and application domains, ranging from image classification and medical segmentation to Transformer-based content generation, demonstrating competitive performance. By extending optical computing to universal, full-range operators across diverse model architectures, DUET provides a viable pathway toward general-purpose optical acceleration for contemporary AI workloads.

## Introduction

In recent years, AI has evolved from a specialized analytical tool into a pervasive technology that increasingly shapes everyday life and industrial systems.[1][2] This transformation has been driven by sustained advances in machine learning and the emergence of foundation models capable of extracting and reasoning over complex multimodal information from the real world.[3][5] Yet these advances have come at the cost of escalating computational demands and energy consumption,[6][7] which now constrain continued scaling in both model size and deployment. This bottleneck is particularly acute during inference,[3][8] where the need for high-throughput service already imposes a substantial computational load and is further aggravated by reasoning-oriented architectures that deliberately increase test-time computation to improve performance.[9] For example, even the relatively compact Llama 2-7B model generates only about 32 tokens per second when executed on an NVIDIA A100 GPU with a thermal design power (TDP) of 250 W.[10][12] Meanwhile, the historical performance gains of electronic processors no longer translate straightforwardly into sustained AI scaling.[13][14] Thermal dissipation limits,[14][15] quantum effects at advanced nanometer technology nodes,[16] and the growing mismatch between computational throughput and electrical data-movement bandwidth have collectively shifted the system bottleneck beyond transistor density alone.[17] Conventional electronic architectures are therefore increasingly unable to sustain the continued growth of contemporary AI workloads, motivating the exploration of alternative computing paradigms.[14][18]

Among emerging alternatives, optical computing is a particularly promising platform for AI acceleration, offering high bandwidth, low latency, low energy dissipation, and intrinsic multiplexing capabilities that are well suited to compute-intensive workloads.[18][19] Recent advances in CMOS-compatible integrated photonics have brought optical computing capabilities to the chip scale, enabling photonic integrated circuits (PICs) that can implement matrix multiplication and foundational neural-network operations.[20][23] These demonstrations have underscored the potential of PIC-based ONNs to achieve high compute density and energy-efficient inference. Yet their scalability remains limited, and most reported systems have therefore been confined to relatively small models and elementary classification or recognition tasks. A primary obstacle is the large spatial footprint of photonic devices, such as modulators, interferometric elements, and photodetectors, which typically occupy micrometer- to millimeter-scale areas and thus constrain the density of optical neurons that can be integrated on a single chip. Additionally, the peripheral electronic circuitry required for electro-optic (E-O) modulation, memory access, and analog-digital interfacing introduces substantial power and latency overheads that erode the intrinsic advantages of optical computing. To mitigate these bottlenecks and unleash the potential of photonic accelerators, recent work has pursued innovations across various dimensions of the technology stack. At the device level, compact optical neurons, exemplified by landmark on-chip diffractive elements and multimode interference (MMI) structures,[24][29] compress optical computation into ultra-dense physical footprints. In parallel, circuit- and architecture-level approaches improve hardware efficiency by restricting the accessible weight space or imposing hardware-aware model compression,[29][31] thereby reducing the required number of photonic elements and active E-O interfaces for a given task. However, both strategies generally trade scalability against reconfigurability or generality, making arbitrary matrix transformations difficult to implement. For instance, diffractive and MMI-based structures are typically passive with weights fixed upon fabrication.[24][27] Although hybrid schemes incorporating localized index tuning or cascaded auxiliary active stages can recover partial reconfigurability, they often require additional optoelectronic control, elaborate calibration, or iterative approximation of target matrices, thereby increasing system complexity and processing latency.[25][26][28] These compromises may be tolerable for conventional weight-stationary models, but they are especially restrictive for contemporary AI workloads, particularly attention mechanisms, whose core computations are dominated by multiplications between dynamic, input-dependent, full-range matrices throughout inference.[3] What remains missing is a hardware-efficient, general-purpose photonic computing primitive that combines compactness with full reconfigurability and supports arbitrary matrix operations across diverse AI workloads.

To bridge this gap, we introduce DUET, a unified photonic computing paradigm built around the VODIC that jointly processes a pair of dynamic, signed vectors within a single interferometric unit. At the device level, VODIC provides a full-range linear encoding interface that natively accommodates arbitrary signed operands, thereby avoiding sign-dependent path splitting, nonlinear remapping, and other preprocessing steps that would otherwise increase latency, memory overhead, and hardware complexity. By applying the corresponding elements of a vector pair to the two terminals of a shared modulation stage, the operand pair is encoded as phase shifts. Intrinsic phase accumulation along the interferometer arms, followed by balanced photodetection, enables compact one-shot multiply-accumulate (MAC) operations directly in the physical domain. Building on this mechanism, DUET functions as a general-purpose photonic tensor core (PTC) that natively accelerates a broad class of operators central to modern AI, including fully connected layers, convolutional layers, and the dynamic, input-dependent matrix multiplications used in attention-based models. To compensate for inevitable on-chip non-idealities, we further develop a HAT strategy grounded in experimental characterization, embedding measured on-chip behavior and device priors into training to improve inference robustness. In this work, we experimentally validate DUET across diverse architectures and application domains, including CNN-based classification, U-Net-style medical image segmentation, and Transformer-based content generation, while achieving competitive task performance and stable inference. Performance analysis indicates that, with appropriate scaling, the proposed design achieves a computational density of 6.01 tera-operations per second (TOPS) per $mm^2$, with power efficiencies of 9.52 TOPS/W and 4.60 TOPS/W for weight-stationary and fully dynamic workloads, respectively. These results demonstrate DUET to be a compact, fully reconfigurable PTC and offer a viable pathway toward general-purpose optical acceleration for contemporary AI workloads.

## Operation Principle

### VODIC for hardware-efficient tensor operation

As the fundamental building block of DUET, the VODIC moves beyond conventional operand encoding, where each operand is typically assigned to an individual modulator, by using differential interferometric signaling to support hardware-efficient tensor operations. Its dual-path interferometric topology integrates multiple independently driven phase shifters, allowing a single cell to process two arbitrary length-$k$ vectors $\boldsymbol{a}, \boldsymbol{b} \in \mathbb{R}^k$. To synthesize a linear dot-product primitive, the VODIC modulates two parallel branches with the sum and difference of the corresponding input-vector components, i.e., $(a_i + b_i)$ and $(a_i - b_i)$ as shown in Fig. 1b. Here, a quadratic mapping between the drive signal and the induced phase shift is required, which can be realized through modulation mechanisms such as thermo-optic actuation, electric-field-induced Kerr effect, or engineered carrier-based phase shifters (see details in Methods and Supplementary Note 7).[32]-[35] With the interferometers biased within the quasi-linear region of their sinusoidal transfer response, the phase modulation is converted into optical intensities proportional to the accumulated squared drive signals. Balanced photodetection then provides a differential readout, cancelling the shared quadratic terms and isolating the desired cross term, as expressed by:

$$
\begin{gathered}
I^{+} \approx \alpha \sum_{i=1}^{k} \left(a_i + b_i\right)^2 + I_0 \, ; \quad I^{-} \approx \alpha \sum_{i=1}^{k} \left(a_i - b_i\right)^2 + I_0 \\
I_{\text{diff}} = I^{+} - I^{-} \approx 4\alpha \sum_{i=1}^{k} a_i b_i \propto \mathbf{a} \cdot \mathbf{b}
\end{gathered}
\tag{1}
$$

Here, $I^+$ and $I^-$ denote the photocurrents from the sum and difference branches, respectively. $I_0$ is the bias photocurrent at the calibrated quasi-linear operating point, and $\alpha$ represents the corresponding modulation coefficient.

This design provides two key advantages over conventional operand-encoding mechanisms in ONNs. First, the VODIC realizes analog MAC operations through a compact differential interferometric cell that combines product extraction and phase-domain accumulation. Each phase-shifting segment receives two signed operands at its two terminals, producing a fixed dual-branch product-extraction response that directly avoids sign-dependent path splitting, nonlinear remapping, and the associated preprocessing-induced latency, memory overhead, and hardware complexity. Additionally, cascading $k$ independently driven segments within the same cell then accumulates these product contributions through intrinsic phase accumulation, enabling physical-domain MAC with high compute density. Second, the dual-branch PIC topology allows half of the drive signals to be shared within the cell, while the complementary polarity of the other operand can be generated electronically with minimal overhead,[36][37] further reducing routing and hardware complexity compared with conventional layouts based on discrete single-operand modulators.

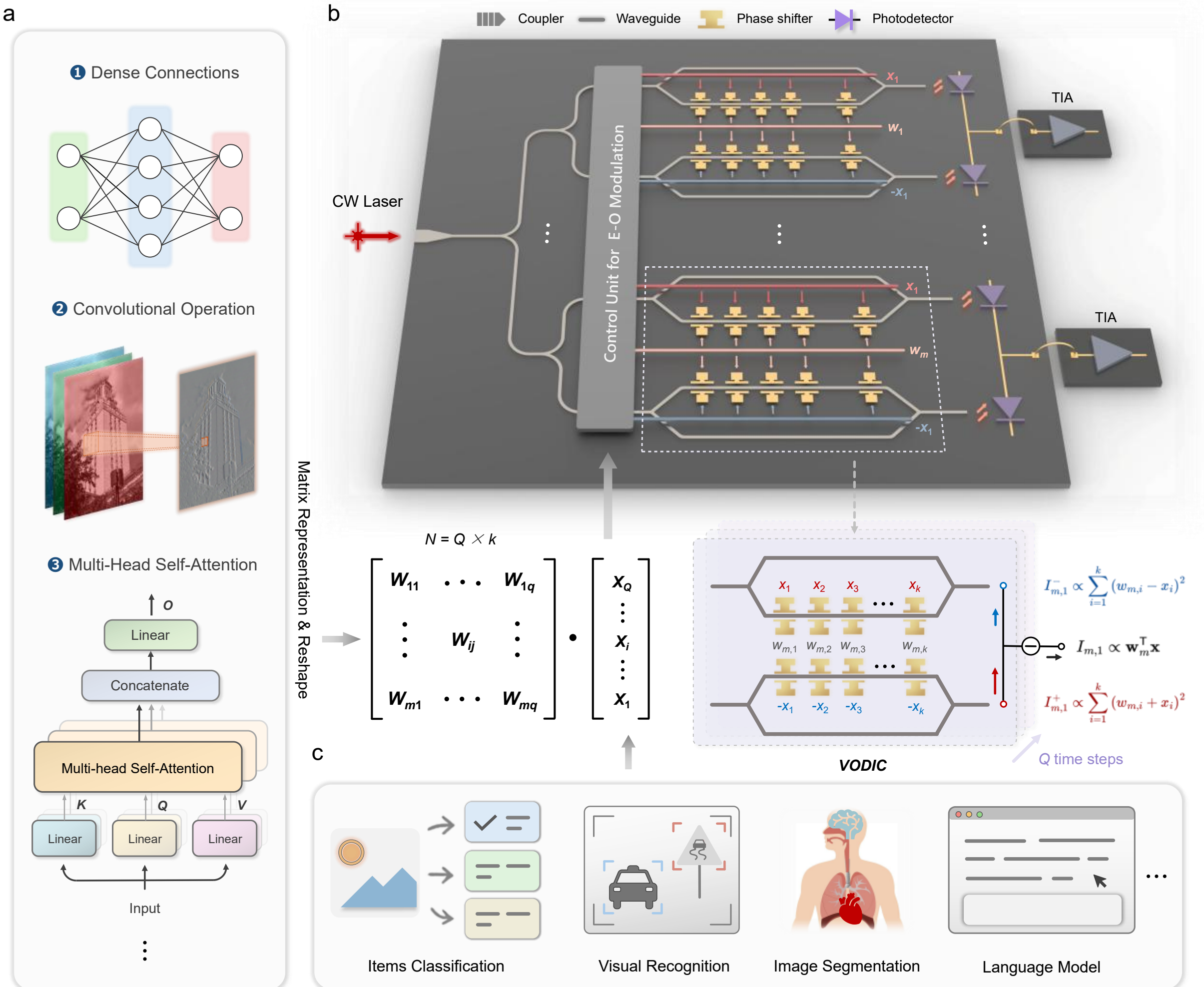

**Fig. 1 Overview of the DUET architecture and implementation of VODIC. a,** DUET supports a broad range of primitives underlying modern AI workloads, including fully connected layers, convolutions, and attention mechanisms. **b,** Schematic of the DUET engine, in which a VODIC array is organized for parallel tensor computation. Although illustrated for matrix-vector multiplication (MVM) between a weight matrix $\boldsymbol{W}$ and an input vector $\boldsymbol{X}$, the architecture supports multiplication between general tensor operands with arbitrary dimensions through vectorized partitioning. **c,** Representative AI workloads and data modalities are reformulated as tensor operations, partitioned into vectorized blocks, and mapped onto DUET for efficient optical AI acceleration.

## DUET architecture for universal AI workloads

Modern AI workloads share a common computational substrate rooted in linear algebra primitives. Across architectures including fully connected, convolutional, recurrent, and attention-based models, the dominant computations can often be expressed as MVMs or general matrix multiplications (GEMMs) through appropriate matrixization techniques (Fig. 1a). For example, convolutional layers commonly use the "*im2col*" operation to unfold kernels and input feature maps into matrices, thereby mapping tensor convolutions onto matrix multiplication.[38][39] Similarly, token sequences in language models are projected into dynamic query ($\boldsymbol{Q}$), key ($\boldsymbol{K}$), and value ($\boldsymbol{V}$) matrices,[3] allowing the core self-attention operations to be expressed as standard GEMMs. This shared algebraic structure provides a unified interface between diverse AI algorithms and the underlying photonic hardware (Fig. 1c).

Building on this unified mathematical foundation, DUET uses a vectorized partitioning strategy to map general tensor operations onto the hardware, as shown in Fig. 1. Specifically, considering a representative operation where an $M \times N$ weight matrix $\boldsymbol{W}$ multiplies an input vector $\boldsymbol{X}$, the matrix is partitioned column-wise into blocks of width $k$, aligned with the number of discrete phase-shifting elements within a VODIC, yielding a total of $Q = \lceil N/k \rceil$ blocks. In this scheme, spatial multiplexing can drive the concurrent processing across both the row and block-level column dimensions, while the full dot product across the column dimension can also be reconstructed through physically accumulating the partial block outputs via photocurrent integration over the $Q$ time steps. This strategy provides a scalable framework for processing tensors of arbitrary dimensions while offering two system-level advantages over conventional ONN architectures. First, beyond the inter-cell signal sharing described above, DUET enables extensive intra-cell sharing of drive signals. This reduces the burden on peripheral electrical circuits and simplifies routing, which is a major bottleneck when scaling ONNs to large tensor dimensions. Second, because the VODIC provides a calibrated linear interface at the cell level, DUET can be tiled into larger arrays without requiring array-level look-up tables (LUTs) correction or operand-dependent remapping. This simplifies peripheral control and supports a unified hardware mapping for signed full-range GEMMs in both weight-stationary and fully dynamic workloads. Such capability is particularly important for Transformer-based language models, where dynamic GEMMs dominate the computational workload and require massive throughput, thereby supporting efficient acceleration of generative AI tasks.

# Results

## Physical implementation and system characterization

To experimentally validate the proposed architecture, we design and tape out a DUET prototype based on VODIC cells, each integrating four modulation sections. For yield and cost considerations, the prototype uses standard thermo-optic modulators from the foundry Process Design Kit (PDK), ensuring consistent device specifications across the chip. The optical micrograph of the fabricated VODIC is shown in Fig. 2a, with magnified insets highlighting the key functional components, while Fig. 2b shows the fully packaged photonic chip. During operation, a continuous-wave (CW) optical signal is coupled into the chip through an edge coupler and distributed by on-chip splitters to individual interferometric cells.

In this work, we apply a systematic two-step calibration protocol to all cells. First, we characterize the phase-shifter responses in both the operand-encoding arms and the bias arm. By sweeping the electrical drive voltage and measuring the optical transmission, we fit the transmission data to a physical interferometer model and extract the tuning efficiency of each phase shifter. The second step aims to identify a high-fidelity linear operating regime within the sinusoidal transfer function. Based on the fitted parameters, we isolate the transmission segment with the highest linearity and bias the device at its lower bound (marked by the grey point in Fig. 2c-e, see more details of calibration in Supplementary Note 1). With a user-defined maximum error threshold of 1.5% relative to an ideal linear response, this procedure yields a usable modulation range of approximately 1.52 radians as illustrated in Fig. 2d. We then extend the same linearization strategy to all on-chip VODICs to ensure consistent modulation behavior. To verify the calibrated system, we implement functional tests using random length-4 input vectors. The differential photocurrents are amplified by transimpedance amplifiers (TIAs), recorded by an oscilloscope, and compared with the theoretical ground truth, defined as the squared $\ell_2$-norm of the input vector ($||\boldsymbol{x}||^2$). As illustrated in Fig. 2e, the measured output closely follows the ideal reference, with a normalized root-mean-square error (NRMSE) of $1.09\times10^{-2}$, confirming the accuracy of the calibrated linear mapping.

Beyond establishing a linear operand-mapping interface, the calibration protocol also provides quantitative characterization of device behavior, which is essential for bridging off-chip training and on-chip measured inference. As an analog computing platform, DUET inevitably exhibits hardware-induced nonidealities, including fabrication variations, noise, and residual nonlinearities within the calibrated operating regime. Consequently, direct deployment of "native" digital models can lead to error accumulation across cascaded layers and substantial performance degradation. To mitigate these physical constraints, we implement a HAT strategy based on experimental characterization. Specifically, we perform an exhaustive sweep of random input vectors to characterize the response statistics of each VODIC and model both static deviations and intensity-dependent heteroscedastic noise, with details provided in Supplementary Note 2. The resulting physical priors are incorporated into a differentiable surrogate model with learnable step-size quantization. To accommodate the finite dynamic range of the modulators, the framework introduces an adaptive scale factor optimized during training, ensuring that operands are mapped effectively within the available drive range while minimizing quantization-induced precision loss. The overall procedure of the HAT framework is summarized in Fig. 2f, with additional details provided in Supplementary Note 2. Notably, this surrogate model encapsulates the dominant physical non-idealities while maintaining full differentiability for end-to-end gradient backpropagation. It further abstracts the hardware response into fundamental computational wrappers, such as linear projections, convolutions, and attention operations, which serve as universal building blocks for constructing modern AI architectures. Compared with in-situ training schemes that update parameters based on direct hardware responses,[40][42] HAT circumvents the requirement for auxiliary monitoring circuitry or real-time feedback loops. By eliminating the substantial time penalties and hardware complexity, this streamlined approach supports efficient scaling to large architectures and provides a practical pathway toward robust, general-purpose ONN deployment.

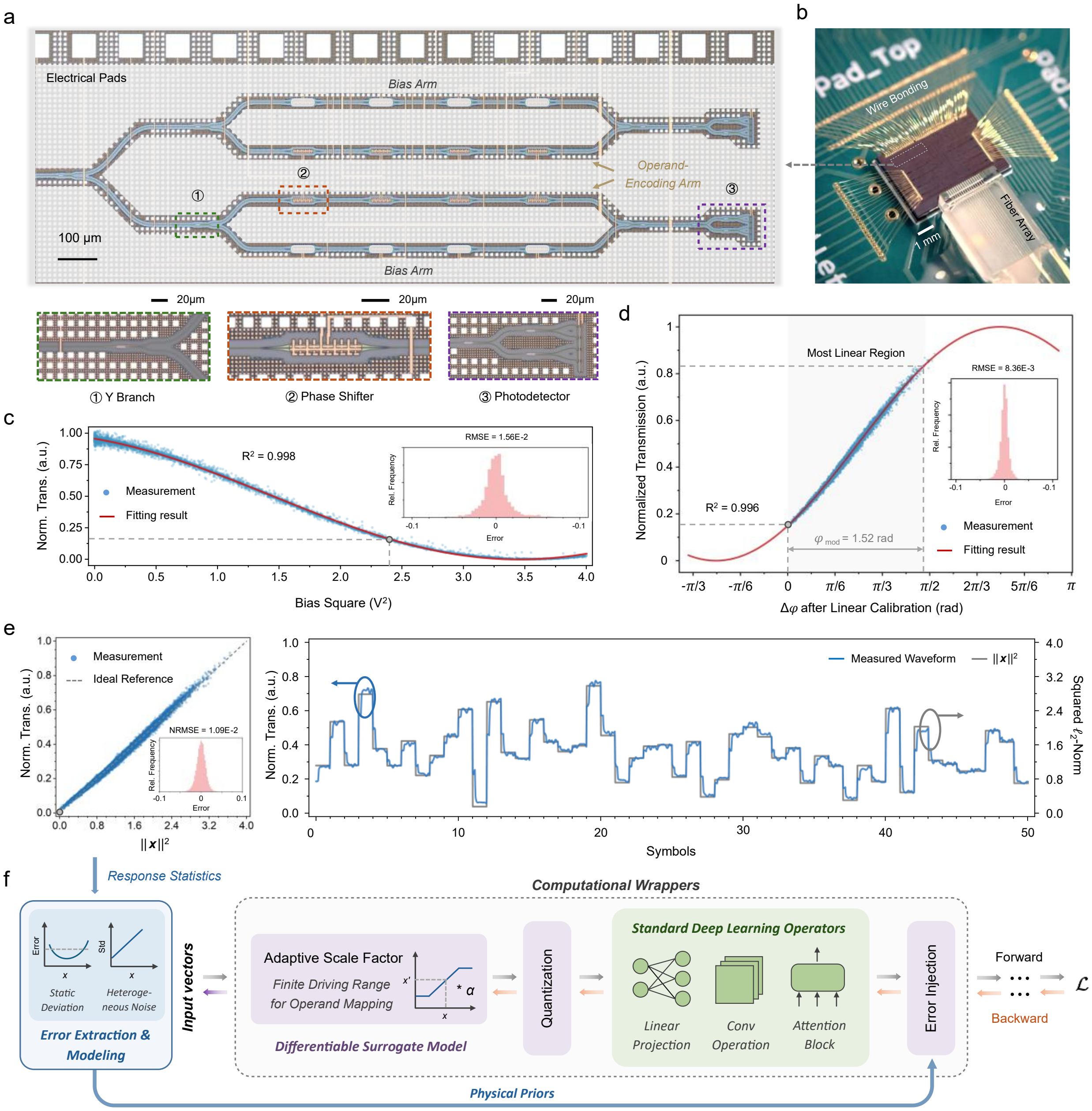

**Fig. 2 Schematic of DUET and system characterization. a,** Layout of a single VODIC, with key functional components highlighted in magnified optical micrographs. **b,** Electrical and optical package for the fabricated DUET prototype. **c & d,** Calibration of the VODIC bias arm and operand-encoding arm, respectively. The measured transmission responses are fitted to physical models, with the insets showing the corresponding residual-error distributions. **e,** Experimental verification of linear operand mapping. Measurements are performed at a modulation frequency of 12.5 kHz, and the oscilloscope samples the output waveform at 500 kHz, corresponding to 40 data points per symbol. **f,** Workflow of the HAT strategy.

## On-chip image processing

To further evaluate DUET and quantify its on-chip MVM accuracy in a practical workload, we implement on-chip image processing with convolutional kernels that extract salient spatial features from input images. As mentioned above, the dominant computations in modern DNNs can be reformulated as one or more MVMs after tensorized rearrangement. Here, we use the standard "*im2col*" transformation for convolution operations.[38][39] Specifically, an $r \times r$ window slides across the image with a unit stride, and each local patch is flattened into a length-$r^2$ vector, and the resulting vectors are stacked to construct a dense operand matrix. The convolution kernels are likewise flattened and assembled into a weight matrix, where each row corresponds to a vectorized kernel for one output channel. Through this reformulation, convolution is reduced to a matrix multiplication natively supported by DUET, followed by reshaping the output back to the spatial grid.

We experimentally validate this operation using an image of the *University of Texas at Austin Main Building* (UT Tower) convolved with multiple convolutional kernels. Because the same spatial two-dimensional kernel is applied to all RGB channels, the input image is reorganized into a unified input matrix with dimensions $r^2 \times 3(H - r + 1)\cdot(W - r + 1)$, where $r = 3$, $H = 336$, and $W = 252$. The host PC vectorizes these operands to align with the VODIC array dimensions before transmitting them to the FPGA-based control unit (Fig. 3a). The data streams are then delivered to the on-chip actuators through DACs with a time interval of $\tau = 80$ μs, corresponding to a symbol rate of 12.5 kbaud. Fig. 3b compares the on-chip convolution results with digital ground truths, while Fig. 3c illustrates the measured waveforms for the Sobel edge-detection kernel against ideal theoretical references (see Supplementary Note 3 for results of all kernels). In these traces, each time slot corresponds to the accumulated output from dot products computed over length-4 subvectors. The reconstructed result achieves an NRMSE of $1.85\times10^{-2}$, demonstrating that DUET can execute large-scale vectorized convolution operations with high fidelity.

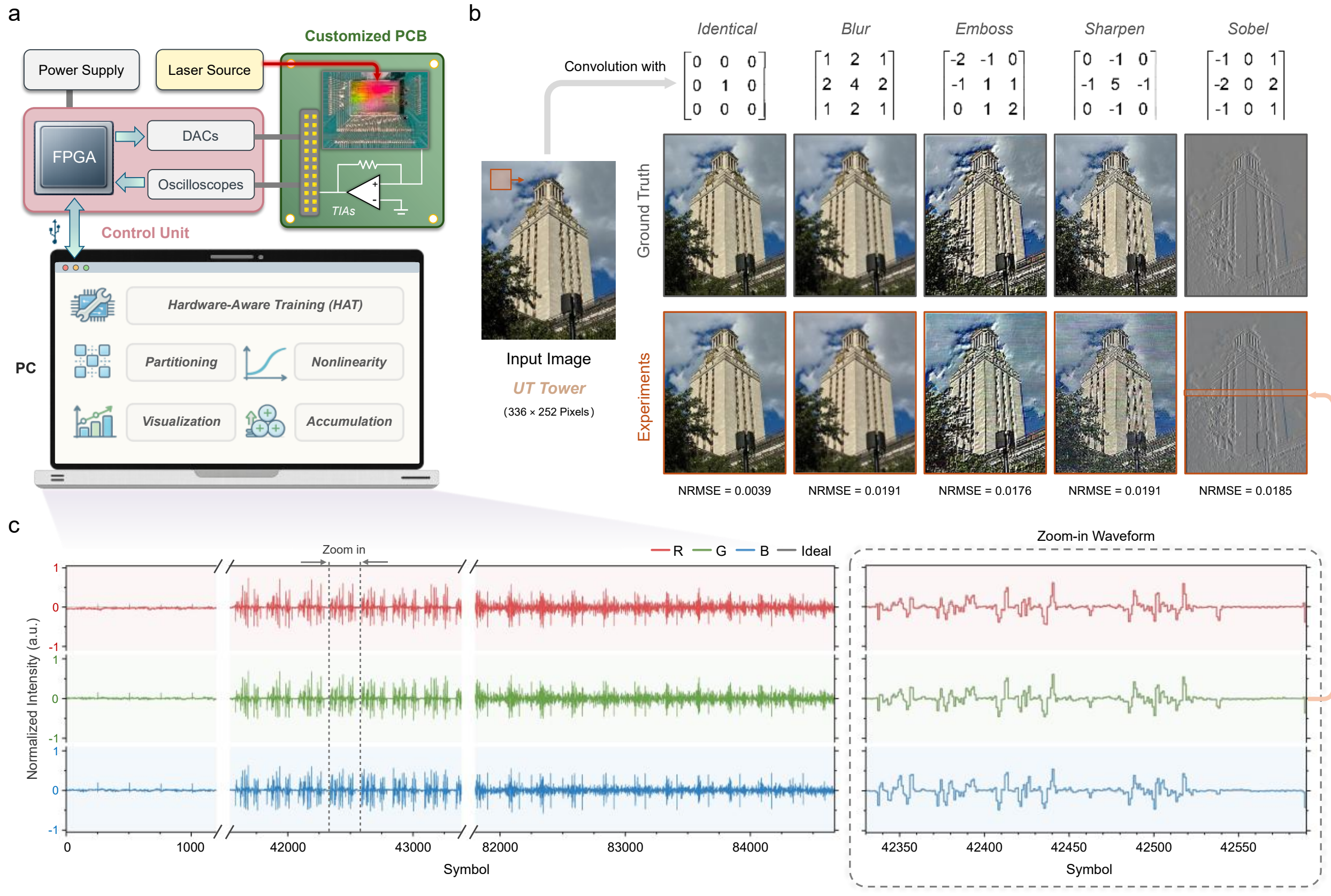

**Fig. 3 Experimental demonstration of on-chip convolutional image processing. a,** Schematic of the experimental setup and test flow, in which input operands are vectorized on the host PC and streamed to DUET through the FPGA-based control unit and multi-channel DACs. **b,** Comparison between digital ground truths and on-chip generated feature maps for the UT Tower image processed by five convolutional kernels: identity, blur, emboss, sharpen, and vertical Sobel. The input photograph is author-acquired. **c,** Representative measured temporal waveforms for the Sobel edge-detection task across the RGB channels.

## DUET-based ONNs on versatile vision workloads

To evaluate the versatility of the proposed architecture across diverse machine-learning tasks, we map the core computational operations of several representative AI models onto the DUET platform. We first deploy lightweight CNNs and assess their performance on two standard vision benchmarks: Fashion-MNIST and the German Traffic Sign Recognition Benchmark (GTSRB), which comprises 43 categories with varying real-world complexity.[43] Both tasks employ a CNN backbone where the DUET chip implements the core convolutional and linear layers, while the remaining operations, including normalization, nonlinear activation, and pooling, are executed in a conventional digital pipeline (Fig. 4a). Depending on task complexity, the deployed models use two convolutional layers for Fashion-MNIST and three convolutional layers for GTSRB.

In these experiments, both weights and input activations are quantized to signed 5-bit resolution. Under this condition, the DUET chip achieves experimental classification accuracies of 90.24% on Fashion-MNIST and 93.71% on GTSRB. The confusion matrix and class-wise accuracy analysis show competitive performance across most categories, indicating that the optical convolutional and linear layers preserve discriminative feature-extraction capability in vision workloads (Fig. 4b,c).

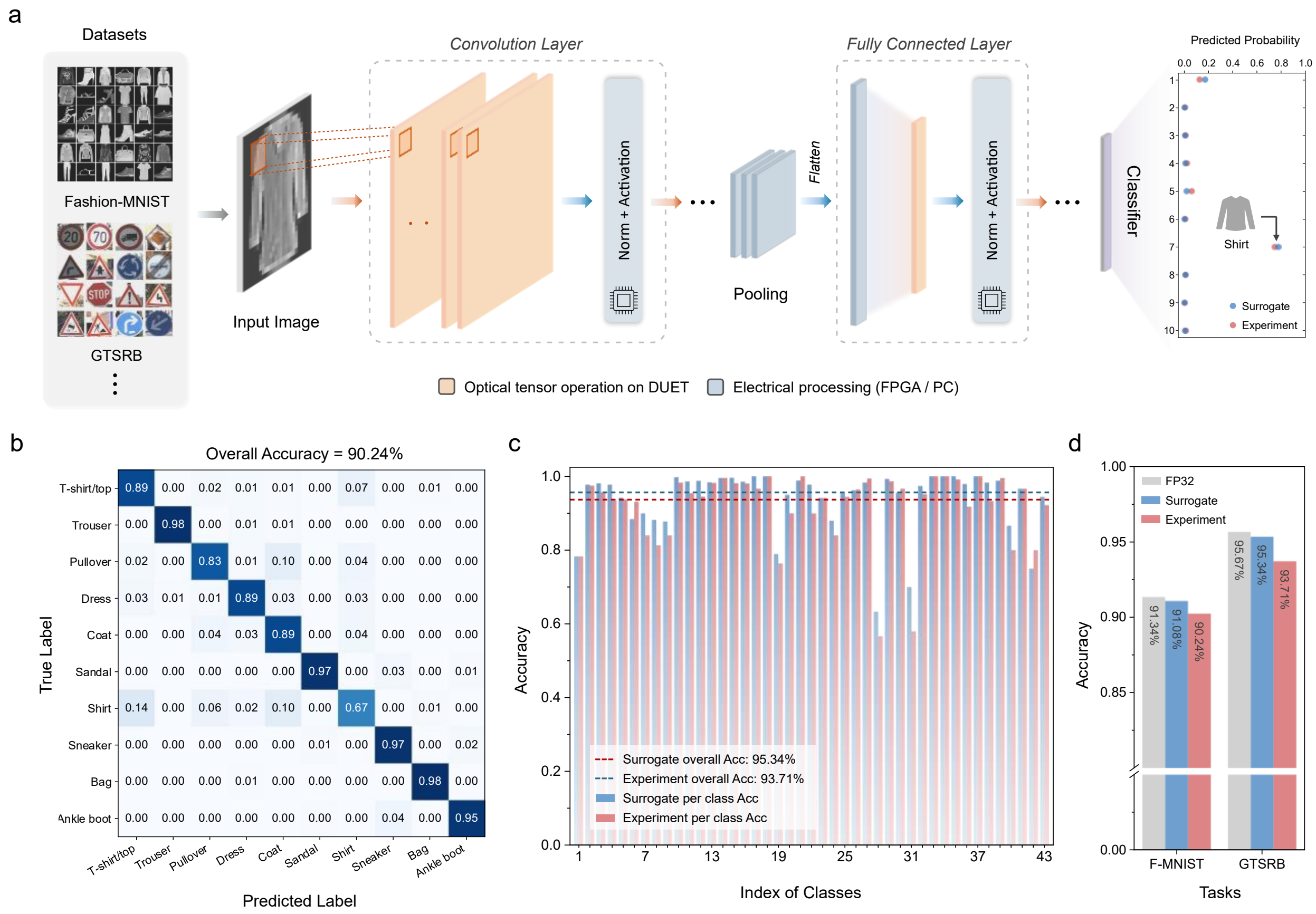

**Fig. 4 Evaluation of the DUET architecture on vision classification tasks. a,** Schematic of the DUET-based CNN inference pipeline, including representative classifier output intensities for a “Shirt” input sample, comparing experimental measurements with surrogate-model predictions. Detailed descriptions of the DUET-based ONN implementation are provided in Supplementary Note 4. **b,** Confusion matrix for the 10-class Fashion-MNIST dataset. **c,** Per-class accuracy for the 43-category GTSRB dataset, comparing surrogate-model predictions with experimental results. Dashed lines denote the corresponding overall accuracies. Because GTSRB has an inherently class-imbalanced sample distribution, the unweighted macro-averaged accuracy of on-chip inference is 91.71%. **d,** Accuracy summary across datasets, comparing full-precision digital baselines using 32-bit floating-point precision (FP32), surrogate-model predictions, and experimental results under identical network configurations.

The observed misclassifications are mainly concentrated in a small subset of visually similar classes (e.g., "Shirt" versus "T-shirt/top" in Fashion-MNIST), aligning with the digital baseline and indicating that these errors primarily arise from intrinsic class-level ambiguity rather than hardware-induced degradation. Importantly, this performance is enabled by the HAT framework discussed in the preceding section. As shown in Fig. 4d, the discrepancy between on-chip experimental results and HAT-based surrogate-model predictions is limited to 0.84 and 1.64 percentage points for Fashion-MNIST and GTSRB, respectively. Compared with full-precision digital baselines using the same CNN configurations, the physical hardware implementation maintains an accuracy reduction of only 1 to 2 percentage points.

Beyond standard image classification, we further evaluate DUET on a dense prediction task that involves a more complex model structure and a spatially resolved output modality. Specifically, we target medical image segmentation using the Brain Tumor Segmentation (BraTS) benchmark,[44][46] where the network delineates tumor subregions from multimodal magnetic resonance imaging (MRI) scans and produces voxel-level anatomical segmentation maps. For this task, we implement a 2.5D U-Net,[47][48] which adopts a standard encoder-decoder architecture with skip connections to aggregate contextual features while progressively restoring spatial resolution. To balance segmentation fidelity and computational overhead, the input pipeline processes three consecutive slices from each modality, and the encoder uses a ResNet-50 backbone (Fig. 5a).[49] Under the hybrid deployment scheme, the computationally intensive 3×3 convolutional layers within the encoder blocks are mapped onto DUET (see details in Supplementary Note 5). The remaining operations, including the 1×1 convolutions for dimensionality reduction and expansion and the decoder stages, are executed in the digital domain. This architectural design and partial physical deployment strategy are jointly optimized to balance system throughput, training stability, and practical end-to-end implementation.

In this task, segmentation performance is evaluated by the Dice coefficient, which quantifies the spatial overlap between the predicted masks and the annotated ground truth. On a test cohort of 70 patients, the surrogate model achieves average Dice scores of 0.761, 0.671, and 0.747 for the whole tumor (WT), tumor core (TC), and enhancing tumor (ET) regions, respectively, showing performance comparable to the FP32 digital baseline. Because the Dice coefficient, by definition, is highly sensitive to small spatial deviations when the target region is small, we compute this metric only for slices in which the ground-truth target region exceeds 25 pixels. Even with this threshold, small or fragmented tumor structures can produce occasional low-Dice outliers, causing the arithmetic mean to fall below the median, as reflected in the box plots (Fig. 5b). Fig. 5c demonstrates the reconstructed tumor segmentations based on on-chip measurements across four representative patient cases. The predicted subregion masks closely align with radiologist annotations in both spatial localization and morphology. Fig. 5d further presents a 3D volumetric rendering for Case IV, comparing the experimentally reconstructed tumor volume with the ground truth across 155 contiguous slices. Full animations for all evaluated cases are provided in the Supplementary Materials.

## Execution of Transformer-based language model

Beyond foundational convolutional and linear operators, we next examine DUET in the context of generative AI, focusing on Transformer-based language models. Transformers underpin a broad range of contemporary generative AI workloads,[50][51] and their computational profile is dominated by self-attention, [3][52] which dynamically couples tokens through data-dependent interactions rather than fixed weights.

However, mapping self-attention onto conventional ONN architectures presents several practical challenges.[18] First, unlike weight-stationary architectures, where one operand (the weight) remains static after training, self-attention requires multiplications between two dynamic, full-range matrices that change throughout inference, including the query-key product $\boldsymbol{QK}^T$ and the subsequent attention-value multiplication. In conventional ONNs, encoding two dynamic, arbitrary matrices generally requires at least two independent sets of modulators. In addition, signed operands often require coherent phase modulation or decomposition into positive and negative components across differential paths, which can introduce up to a fourfold increase in hardware overhead or processing time and further complicate control. Second, Transformers are often operated in highly demanding, throughput-critical regimes, such as large language models (LLMs) and vision transformers (ViTs),[50][51][53] where new input-dependent matrices must be mapped to hardware at each token step. In conventional PTCs, this mapping can require continuous digital preprocessing,[20] such as matrix factorization for interferometer meshes, as well as memory-intensive look-up operations that compensate for modulator nonidealities at the E-O interface. When these transformations are placed in the active token-rate computation path, they can introduce a substantial digital bottleneck in end-to-end latency and energy consumption.

DUET circumvents these fundamental limitations through its device-level encoding mechanism and unique dataflow. The architecture directly encodes two dynamic operands within a shared actuator stage, while its compact layout and spatial data-sharing capability support high computing density without prohibitive area scaling. The calibrated linear modulation interface further eliminates the need for dynamic operand preprocessing and simplifies the aggregation of partial

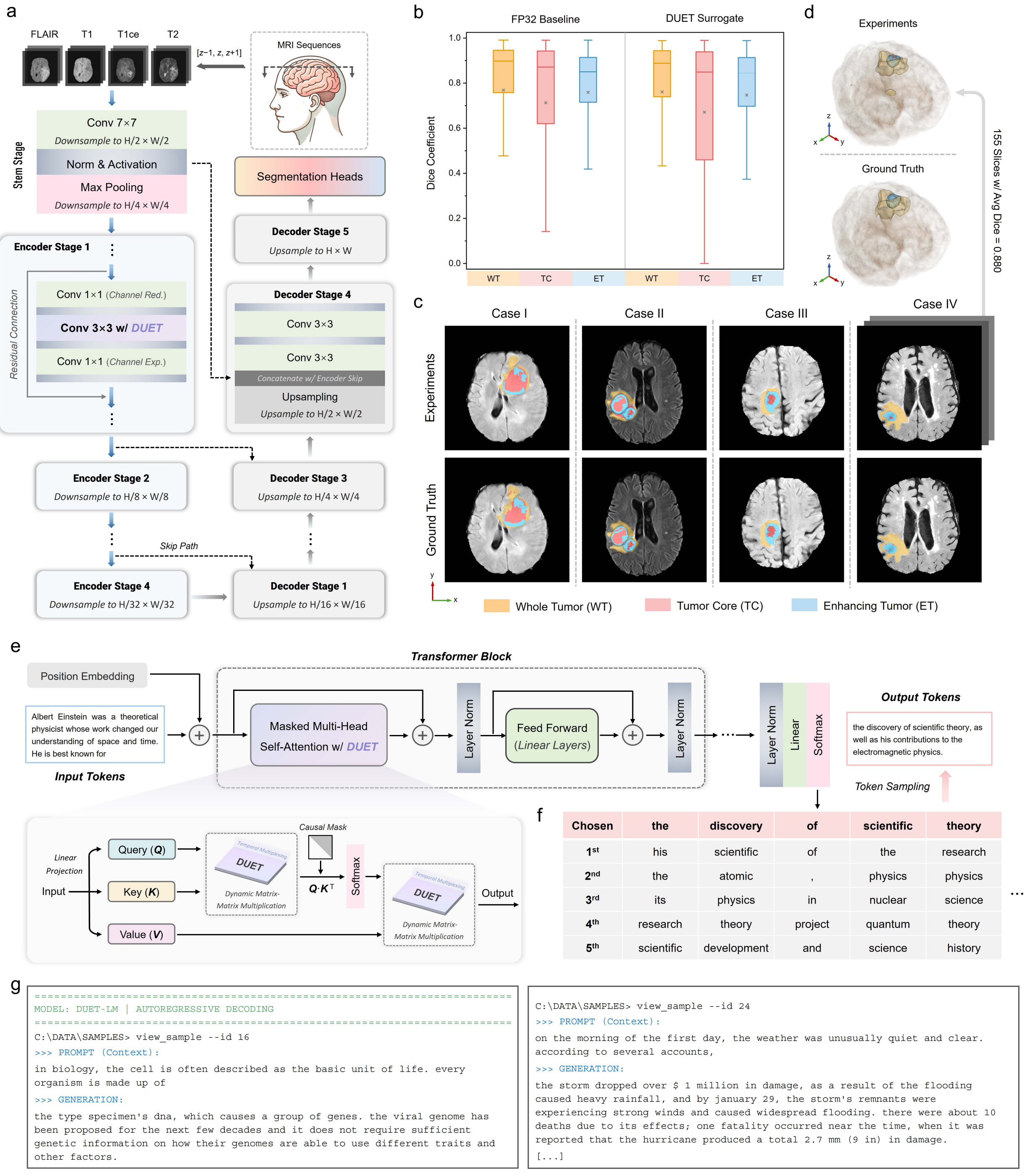

**Fig. 5 Extending DUET to medical segmentation and dynamic self-attention. a,** U-Net architecture configured for brain tumor segmentation. **b**, Box-plot comparison of segmentation performance across the three tumor subregions. **c & d,** Representative segmentation results from selected cases, shown as slice-wise overlays and three-dimensional volumetric renderings, respectively, comparing DUET-reconstructed predictions with ground-truth annotations. **e**, Schematic and execution flow of a decoder-only Transformer with DUET-based self-attention. **f,** Representative decoding snapshot showing the top-five token candidates during autoregressive generation. **g,** Sample prompts and corresponding outputs across representative topical domains.

optical outputs during large-scale tensor operations. Although the proof-of-concept prototype demonstrated in this work uses thermo-optic tuning with limited modulation speed, the architecture is compatible with high-speed E-O modulators operating at tens of GHz. Detailed device-level implementations and performance projections are provided in Methods and Supplementary Note 7.[35][54]

Following similar design principles to balance throughput, robustness, and hardware constraints, we implement a customized lightweight autoregressive Transformer with 25.8 million parameters on the WikiText corpus for next-token prediction. Because this evaluation specifically targets dynamic attention execution, the on-chip deployment focuses on the multi-head self-attention module. Specifically, within a standard Transformer block (Fig. 5e), the two core dynamic matrix-matrix multiplications, $\boldsymbol{QK}^T$ and $\boldsymbol{PV}$, where $\boldsymbol{P} = \mathrm{softmax}(\boldsymbol{QK}^T/\sqrt{d_k})$, are executed on DUET, whereas the remaining operations are handled by digital processors. During generation, the input prompt is first vectorized through token embeddings and then propagated through the network to produce a probability distribution over the vocabulary for the next token. Fig. 5f illustrates a snapshot of the decoding process, including the five highest-probability candidates at each generation step. Because standard stochastic sampling strategies (e.g., temperature scaling; see Supplementary Note 6 for details) are applied to maintain generation diversity, the final appended token is drawn probabilistically and may fall outside these top candidates.[50][55][56] Across diverse topical prompts, the generated sequences maintain topical relevance and contextual alignment with the input prompts (Fig. 5g; complete generation results across different prompts are provided in the Appendix of Supporting Information). Although the generated text can exhibit semantic drift, grammatical inconsistencies, and factual hallucinations,[55] such behaviors are expected for a modest-scale model trained on a limited corpus, whereas the overall structural coherence remains intact.[6] Therefore, these results support the feasibility and fidelity of executing dynamic attention mechanisms on the DUET platform.

## Discussion

We assess the system-level potential of the proposed architecture through numerical analyses across multiple benchmark configurations, with details provided in Supplementary Note 8. With appropriate scaling and high-speed modulation, DUET reaches a projected computational density of 6.01 TOPS/mm$^2$. Additionally, overall power efficiency is highly dependent on the operating mode. In the weight-stationary regime, extensive data reuse enables a power efficiency of up to 9.52 TOPS/W, whereas fully dynamic operation sustains 4.60 TOPS/W at an optimal block configuration of $M = 128$ and $k = 24$. Notably, the primary power bottleneck in both regimes stems from the peripheral electronics rather than the photonic core itself. For weight-stationary operation, power consumption is dominated by the receiver front end (ADCs and TIAs), which could be amortized through multi-cycle charge-domain integration with lower-rate digitization, or reduced by aggregating photocurrents column-wise from parallel blocks before amplification. In the fully dynamic regime, DAC power becomes the leading power bottleneck, reflecting the unavoidable cost of refreshing both operand streams at high speed. Unlocking higher system efficiencies will therefore necessitate co-optimizing the E-O interfaces with state-of-the-art, low-energy mixed-signal electronics.

Beyond peripheral power constraints, these metrics reveal a fundamental architectural trade-off. Because DUET relies on electrical rather than optical signal broadcasting, it does not fully exploit the inherent multiplexing advantages of light, and its computing density may not match that of architectures explicitly optimized for massive optical fan-out or compressed weight spaces.[24][31] However, the distinct advantage of DUET lies in its functional versatility for general-purpose inference acceleration, particularly for attention-based workloads. By natively supporting arbitrary dynamic matrix operations, it addresses a gap in existing photonic architectures that rely on static or structured weights. This complementary capability motivates a heterogeneous optical-computing paradigm for future AI hardware, in which highly dynamic workloads are assigned to DUET, whereas compressible or weight-stationary tasks are mapped to specialized high-density PTCs.

In summary, we present DUET, a general-purpose photonic computing paradigm built on differential dual-operand interferometric encoding for full-range dynamic tensor operations. By executing signed tensor operations in the physical domain, the architecture bypasses the decomposition, nonlinear remapping, and auxiliary preprocessing steps, thereby reducing architectural overhead while preserving computational generality. Furthermore, the HAT framework improves the inference robustness against on-chip non-idealities and limited control resolution by embedding experimental device characteristics directly into the training pipeline. Notably, DUET delivers competitive, general-purpose acceleration across complex, heterogeneous AI models, highlighting its versatility as a foundational computing unit for modern AI workloads. Beyond the demonstrated thermo-optic prototype, the underlying operating principle is compatible with high-speed E-O modulation and compact photonic-electronic integration, offering a viable pathway toward high-throughput and massive-scale computing. These results establish DUET as a reconfigurable photonic-electronic primitive for dynamic signed tensor operations, complementing specialized optical accelerators designed for static or structured workloads, bringing light closer to serving as a practical substrate for the next-generation high-performance computing.

## Funding

Multidisciplinary University Research Initiative (MURI) Program (FA9550-17-1-0071); Air Force Office of Scientific Research (AFOSR) (FA9550-23-1-0452).

## Disclosures

The authors declare no conflicts of interest.

## Data availability

Data underlying the results presented in this paper are available from the authors upon reasonable request.

## Supplemental material

See supplementary materials for supporting contents.

## Methods

### Experimental setup

In this work, all devices and instruments used for testing, except for the DUET chip fabricated through *AIM Photonics*, are commercially available components. The PIC layout was developed and verified in *Synopsys Optodesigner*. Electrical control signals were routed to edge pads and wire-bonded to a custom printed circuit board (PCB). These signals were generated by a 40-channel, 14-bit DAC (AD5370, Analog Devices), controlled through the serial peripheral interface (SPI) by an FPGA board (PYNQ-Z2), and subsequently buffered by analog amplifiers (TLE2064, Texas Instruments). Power was supplied by a triple-output DC source (E3630A, Agilent), providing ±12 V for the analog circuitry and a 1 V bias for the photodetector. Optical inputs were provided by tunable continuous-wave laser sources (CoBrite DX4, ID Photonics), with the input polarization adjusted by fiber polarization controllers (OZ Optics). Light was coupled into the chip through edge couplers using a single-mode fiber array (Meisu Technology Co., Ltd.). The output photocurrents were amplified by TIAs (LT1216CN, Analog Devices) and recorded with oscilloscopes (Analog Discovery 2, Digilent). The overall measurement platform was coordinated by the FPGA, while data acquisition was performed through the oscilloscope application programming interface (API).

### Working principle of HAT

To improve robustness against on-chip non-idealities, we propose a physics-guided HAT framework that incorporates experimentally measured device behavior into offline training. For each interferometric cell, the ideal response is defined as $s = \|\boldsymbol{v}\|^2$, where $\boldsymbol{v}$ denotes the encoded input vector. The residual is expressed as $r = y - s$, where $y$ represents the measured output from the device. Based on the calibration data, this residual is decomposed into a deterministic bias term $\mu(s)$ and a stochastic noise term $\varepsilon(s)$ with standard deviation $\sigma(s)$, yielding a surrogate model of the form $y = s + \mu(s) + \varepsilon(s)$. The conditional mean and variance are extracted by binning the measured samples along $s$. During training, the noise term is approximated as signal-dependent Gaussian noise, whereas the static bias is modeled either by polynomial fitting or a one-dimensional LUT constructed from the binned measurements. Because the extracted noise statistics corresponded to a system SNR of approximately 32 dB, equivalent to an effective hardware precision of about 5.0-5.4 bits, 5-bit quantization is adopted as the hardware-matched operating point for subsequent ONN implementations. Detailed derivations and implementation are provided in Supplementary Note 2.

## Implementation of high-speed DUET

To support the high-throughput demands of contemporary AI workloads, DUET must be adapted for high-speed E-O implementations. Carrier-based silicon modulators operating in depletion or accumulation mode provide a practical route toward high-speed implementations of DUET, as such devices are compatible with mature silicon photonic platforms and support bandwidths in the tens-of-GHz regime. Although their voltage-to-phase responses are not strictly quadratic, they can be accurately described by a low-order polynomial within the target operating range. To synthesize the bilinear term required for DUET, we propose a symmetric quaternary differencing scheme. In this approach, four bias-centered composite drives, $V_{\sigma,\tau} = V_b + \alpha(\sigma x + \tau y)$ with $\sigma, \tau$ in {+1, -1}, are applied and combined as $\varphi(V_{++}) - \varphi(V_{+-}) - \varphi(V_{-+}) + \varphi(V_{--})$. This operation cancels the odd-order contributions and isolates a quadratic-dominant response proportional to $xy$, thereby enabling product extraction from carrier-based modulators. Experimental characterization of a depletion-mode Mach-Zehnder interferometer confirmed that the transfer function is well captured by a low-order polynomial, with the first- and second-order terms dominating the response, thereby validating this approach. Relative to the native quadratic thermo-optic prototype, this general implementation requires an additional constant-factor resource trade-off, as the four-point extraction can be realized either by two temporal differential evaluations or by spatial interleaving of the corresponding drive/readout lanes. Importantly, this cost is independent of operand sign patterns and therefore differs from the sign-decomposition overhead required in conventional intensity-encoded ONNs. Furthermore, this formulation remains applicable even when device sign conventions dictate a strictly negative operating bias. In this case, the common bias either cancels out or is absorbed into the effective even-order coefficients. See Supplementary Note 7 for detailed derivations and demonstrations.

## References

[1] LeCun, Y., Bengio, Y. & Hinton, G. Deep learning. *Nature* **521**, 436–444 (2015).

[2] Jordan, M. I. & Mitchell, T. M. Machine learning: trends, perspectives, and prospects. *Science* **349**, 255–260 (2015).

[3] Vaswani, A. *et al.* Attention is all you need. *Adv. Neural Inf. Process. Syst.* **30**, 5998–6008 (2017).

[4] Radford, A. *et al.* Learning transferable visual models from natural language supervision. In *International Conference on Machine Learning* 8748–8763 (PMLR, 2021).

[5] Brown, T. *et al.* Language models are few-shot learners. *Adv. Neural Inf. Process. Syst.* **33**, 1877–1901 (2020).

[6] Kaplan, J. *et al.* Scaling laws for neural language models. *arXiv preprint* arXiv:2001.08361 (2020).

[7] Patterson, D. *et al.* Carbon emissions and large neural network training. *arXiv preprint* arXiv:2104.10350 (2021).

[8] Thompson, N.C., Greenewald, K., Lee, K. and Manso, G.F., 2020. The computational limits of deep learning. *arXiv preprint* arXiv:2007.05558, *10*(2).

[9] Guo, D., Yang, D., Zhang, H., Song, J., Wang, P., Zhu, Q., Xu, R., Zhang, R., Ma, S., Bi, X. and Zhang, X., 2025. Deepseek-R1: Incentivizing reasoning capability in llms via reinforcement learning. *arXiv preprint* arXiv:2501.12948.

[10] Touvron, H. *et al.* Llama 2: open foundation and fine-tuned chat models. *arXiv* arXiv:2307.09288 (2023).

[11] Dell Technologies. Llama 2 inferencing on a single GPU. *Info Hub* https://infohub.delltechnologies.com/t/llama-2-inferencing-on-a-single-gpu/ (2023). Accessed 1 March 2026.

[12] NVIDIA. NVIDIA A100 Tensor Core GPU datasheet. https://www.nvidia.com/en-us/data-center/a100/. Accessed 1 March 2026.

[13] Waldrop, M. M. More than Moore. *Nature* **530**, 144–148 (2016).

[14] Shalf, J. The future of computing beyond Moore's Law. Philos. *Trans. R. Soc. A* **378**, 20190061 (2020).

[15] Horowitz, M. 1.1 computing's energy problem and what we can do about it. In *2014 IEEE International Solid-State Circuits Conference Digest of Technical Papers* 10–14 (2014).

[16] Fang, F. *et al.* Towards atomic and close-to-atomic scale manufacturing. *Int. J. Extrem. Manuf.* **1**, 012001 (2019).

[17] Reed, D., Gannon, D. & Dongarra, J. Reinventing high performance computing: challenges and opportunities. *arXiv preprint* arXiv:2203.02544 (2022).

[18] Ning, S. *et al.* Photonic-electronic integrated circuits for high-performance computing and AI accelerators. *J. Lightwave Technol.* **42**, 7834–7859 (2024).

[19] Shastri, B. J. *et al.* Photonics for artificial intelligence and neuromorphic computing. *Nat. Photonics* **15**, 102–114 (2021).

[20] Shen, Y. *et al.* Deep learning with coherent nanophotonic circuits. *Nat. Photonics* **11**, 441–446 (2017).

[21] Tait, A. N. *et al.* Neuromorphic photonic networks using silicon photonic weight banks. *Sci. Rep.* **7**, 7430 (2017).

[22] Feldmann, J. *et al.* Parallel convolutional processing using an integrated photonic tensor core. *Nature* **589**, 52–58 (2021).

[23] Xu, X. *et al.* 11 TOPS photonic convolutional accelerator for optical neural networks. *Nature* **589**, 44–51 (2021).

[24] Fu, T. *et al.* Photonic machine learning with on-chip diffractive optics. *Nat. Commun.* **14**, 70 (2023).

[25] Xu, Z. *et al.* Large-scale photonic chiplet Taichi empowers 160-TOPS/W artificial general intelligence. *Science* **384**, 202–209 (2024).

[26] Wang, C., Cheng, Y., Xu, Z., Dai, Q. & Fang, L. Diffractive tensorized unit for million-TOPS general-purpose computing. *Nat. Photonics* **19**, 1078–1087 (2025).

[27] Meng, X. *et al.* Compact optical convolution processing unit based on multimode interference. *Nat. Commun.* **14**,

3000 (2023).

[28] Li, J. *et al.* End-to-end closed-loop optoelectronic computing breaking precision-accuracy coupling. *Adv. Photonics* **8**, 016005 (2026).

[29] Zhu, H. H. *et al.* Space-efficient optical computing with an integrated chip diffractive neural network. *Nat. Commun.* **13**, 1044 (2022).

[30] Feng, C. *et al.* A compact butterfly-style silicon photonic-electronic neural chip for hardware-efficient deep learning. *ACS Photonics* **9**, 3906–3916 (2022).

[31] Ning, S. *et al.* Hardware-efficient photonic tensor core: accelerating deep neural networks with structured compression. *Optica* **12**, 1079–1089 (2025).

[32] Timurdogan, E., Poulton, C. V., Byrd, M. J. & Watts, M. R. Electric field-induced second-order nonlinear optical effects in silicon waveguides. *Nat. Photonics* **11**, 200–206 (2017).

[33] Peltier, J. *et al.* High-speed silicon photonic electro-optic Kerr modulation. *Photonics Res.* **12**, 51–60 (2023).

[34] Xia, P. *et al.* High linearity silicon DC Kerr modulator enhanced by slow light for 112 Gbit/s PAM4 over 2 km single mode fiber transmission. *Opt. Express* **30**, 16996–17007 (2022).

[35] Hiraki, T. *et al.* Heterogeneously integrated III-V/Si MOS capacitor Mach-Zehnder modulator. *Nat. Photonics* **11**, 482–485 (2017).

[36] Berikaa, E., Alam, M. S., Hu, Y., Li, W. & Plant, D. V. C-band 100 Gb/s transmission over 40 km SSMF using a silicon photonic vestigial sideband transmitter based on dual-drive MZM and passive optical delay line. In *Optical Fiber Communication Conference (OFC)*, paper Th3E.7 (2023).

[37] Dong, P., Chen, L. & Chen, Y. K. High-speed low-voltage single-drive push-pull silicon Mach-Zehnder modulators. *Opt. Express* **20**, 6163–6169 (2012).

[38] Chellapilla, K., Puri, S. & Simard, P. High performance convolutional neural networks for document processing. In *Tenth International Workshop on Frontiers in Handwriting Recognition* (2006).

[39] Jia, Y. *et al.* Caffe: convolutional architecture for fast feature embedding. In *Proceedings of the 22nd ACM International Conference on Multimedia* 675–678 (2014).

[40] Xue, Z. *et al.* Fully forward mode training for optical neural networks. *Nature* **632**, 280–286 (2024).

[41] Wu, B. *et al.* Scaling up for end-to-end on-chip photonic neural network inference. *Light Sci. Appl.* **14**, 328 (2025).

[42] Zhao, B. *et al.* In-situ trained microring-based neural networks for scalable and robust photonic computing. *Laser Photonics Rev.* **20**, 2501576 (2025).

[43] Stallkamp, J., Schlipsing, M., Salmen, J. & Igel, C. The German traffic sign recognition benchmark: a multi-class classification competition. In *The 2011 International Joint Conference on Neural Networks* 1453–1460 (2011).

[44] Menze, B. H. *et al.* The multimodal brain tumor image segmentation benchmark (BRATS). *IEEE Trans. Med. Imaging* **34**, 1993–2024 (2014).

[45] Bakas, S. *et al.* Advancing the cancer genome atlas glioma MRI collections with expert segmentation labels and radiomic features. *Sci. Data* **4**, 170117 (2017).

[46] Bakas, S. *et al.* Segmentation labels for the pre-operative scans of the TCGA-GBM collection. *The Cancer Imaging Archive* (2017).

[47] Ronneberger, O., Fischer, P. & Brox, T. U-net: convolutional networks for biomedical image segmentation. In *International Conference on Medical Image Computing and Computer-Assisted Intervention* 234–241 (2015).

[48] Çiçek, Ö., Abdulkadir, A., Lienkamp, S. S., Brox, T. & Ronneberger, O. 3D U-Net: learning dense volumetric segmentation from sparse annotation. In *International Conference on Medical Image Computing and Computer-*

*Assisted Intervention* 424–432 (2016).
[49] He, K., Zhang, X., Ren, S. & Sun, J. Deep residual learning for image recognition. In *Proceedings of the IEEE Conference on Computer Vision and Pattern Recognition* 770–778 (2016).
[50] Radford, A. *et al.* Language models are unsupervised multitask learners. *OpenAI Blog* **1**, 9 (2019).
[51] Devlin, J., Chang, M. W., Lee, K. & Toutanova, K. Bert: pre-training of deep bidirectional transformers for language understanding. In *Proceedings of the 2019 Conference of the North American Chapter of the Association for Computational Linguistics: Human Language Technologies, Volume 1 Long and Short Papers* 4171–4186 (2019).
[52] Bahdanau, D., Cho, K. & Bengio, Y. Neural machine translation by jointly learning to align and translate. *arXiv preprint* arXiv:1409.0473 (2014).
[53] Dosovitskiy, A. *et al.* An image is worth 16x16 words: transformers for image recognition at scale. In *International Conference on Learning Representations* (2021).
[54] Wang, J. *et al.* Optimization and demonstration of a large-bandwidth carrier-depletion silicon optical modulator. *J. Lightwave Technol.* **31**, 4119–4125 (2013).
[55] Holtzman, A., Buys, J., Du, L., Forbes, M. & Choi, Y. The curious case of neural text degeneration. In *International Conference on Learning Representations* (2020).
[56] Fan, A., Lewis, M. & Dauphin, Y. Hierarchical neural story generation. In *Proceedings of the 56th Annual Meeting of the Association for Computational Linguistics Volume 1: Long Papers* 889–898 (2018).